# Efficient computation of demagnetising fields for magnetic multilayers using multilayered convolution

Serban Lepadatu[*]

*Jeremiah Horrocks Institute for Mathematics, Physics and Astronomy, University of Central Lancashire, Preston PR1 2HE, U.K.*

## Abstract

As research into magnetic thin films and spintronics devices is moving from single to multiple magnetic layers, there is a need for micromagnetics modelling tools specifically designed to efficiently handle magnetic multilayers. Here we show an exact method of computing demagnetising fields in magnetic multilayers, which is able to handle layers with arbitrary spacing, arbitrary thicknesses, and arbitrary relative positioning between them without impacting on the computational performance. The multilayered convolution method is a generalisation of the well-known fast Fourier transform-based convolution method used to compute demagnetising fields in a single magnetic body. In typical use cases, such as multilayered stacks used to study skyrmions, we show the multilayered convolution method can be up to 8 times faster, implemented both for central processors and graphics processors, compared to the simple convolution method.

Keywords: micromagnetics, skyrmions, magnetic multilayers, demagnetising field

[*] SLepadatu@uclan.ac.uk

# 1. Introduction

Multilayered magnetic structures are currently at the fore-front of research into spintronics devices, spurred by applications to non-volatile magnetic memories and logic, as well as the fascinating physics of spin transport across multiple ferromagnetic / non-magnetic layer interfaces. In particular skyrmions [1], stabilised at room temperature in ultrathin magnetic layers through the Dzyaloshinskii-Moriya interaction (DMI) [2,3], have shown great promise as information carriers in spintronics devices, utilising the spin-Hall effect to efficiently manipulate them with electrical currents [4]. Skyrmion motion has been observed in magnetic multilayered stacks [5-8], whilst hybrid chiral skyrmions have been studied in magnetic multilayers [9,10]. Racetrack memory devices have also been proposed [11], based on current-induced domain wall motion in multilayered stacks [12,13]. Moreover, magnetic multilayers with surface exchange coupling allow the modification of dipolar interactions in synthetic antiferromagnetic and synthetic ferrimagnetic tracks [14-17], resulting in fast domain wall motion and reduced threshold currents.

Numerical micromagnetics [18] modelling plays a very important part in understanding and analysing experimental results, allowing reproduction of magnetisation dynamics in the presence of magnetic fields, as well as spin torques in multilayers [19]. The magnetostatic interaction, which is an essential part of the micromagnetics model, is particularly difficult to evaluate due to its long-range effect, accounting for the majority of simulation time. The use of magnetic multilayers further significantly complicates this, as the spacing and thicknesses of magnetic layers used in many experimental studies make it difficult to discretise the simulation space, whilst also allowing an efficient simulation. A widely-used approach to calculating the demagnetising fields due to the magnetostatic interaction, is based on finite difference discretisation, and uses fast Fourier transforms (FFT) to evaluate the convolution sum of a demagnetising tensor with the magnetisation distribution [20,21]. A closely related method allows calculation of demagnetising fields from the scalar potential [22,23]. When applied to multilayers, the main difficulty with this approach is the requirement for uniform computational mesh discretisation. This poses a problem for magnetic multilayers, where the layer thicknesses and spacings may not readily lend themselves to a uniform discretisation of the entire space. Other approaches to calculating the demagnetising field are available, including tensor grids [24], as well as finite element /

boundary element methods [25] – for a review of this class of methods see Ref. [26]. The finite difference method with FFT-based convolution remains very popular, owing to better computational performance compared to finite element methods, particularly for rectangular geometries [26]. Finite element methods are more accurate for curved geometries, although staircase corrections can be used in the finite difference formulation to reduce discretisation errors on the demagnetising field [27,28].

Freely available software include OOMMF [29], mumax3 [30], and Fidimag [31], and all compute demagnetising fields using the FFT-based convolution method. Here we introduce a new method specifically designed for magnetic multilayers, which is a generalisation of the FFT-based convolution method, termed multilayered convolution. This method has been implemented in Boris Computational Spintronics [32] both for central processors (CPU) and graphics processors (GPU). Multilayered convolution allows computation of demagnetising fields in multiple layers with arbitrary thicknesses and spacing, without the requirement to uniformly discretise the entire simulation space. In typical use cases we show this method results in computational speeds up to 8 times faster compared to FFT-based convolution with uniform discretisation, whilst still being an exact method.

## 2. Multilayered Convolution

In micromagnetics, for a magnetic body with a discrete distribution of magnetisation values **M** at points in the set $V = \{\mathbf{r}_i;\ i \in P\}$, the demagnetising fields may be obtained as:

$$\mathbf{H}(\mathbf{r}_k) = -\sum_{\mathbf{r}_i \in V} \mathbf{N}(\mathbf{r}_k - \mathbf{r}_i)\mathbf{M}(\mathbf{r}_i), \quad \mathbf{r}_k \in V \tag{1}$$

The demagnetising tensor **N** has the following components, which may be computed using the formulas given in Newell et al. [33]:

$$\mathbf{N} = \begin{pmatrix} N_{xx} & N_{xy} & N_{xz} \\ N_{xy} & N_{yy} & N_{yz} \\ N_{xz} & N_{yz} & N_{zz} \end{pmatrix} \tag{2}$$

For uniform finite difference discretisation, Equation **(**1**)** may be evaluated very efficiently using the convolution theorem [20,21]: the input magnetisation and tensor components are transformed using the discrete Fourier transform (DFT), multiplied point-by-point in the transform space, and the output demagnetising field distribution is obtained by further applying the inverse DFT. Since the demagnetising tensor only depends on the fixed geometry, it can be obtained in kernel form by applying the DFT only once in the initialisation stage.

When we have a collection of magnetic bodies, $\{V_i;\ i = 1, \ldots, n\}$, one approach to calculating the demagnetising field distribution is to simply apply Equation **(**1**)** again by taking the union of these separate magnetic bodies into a single magnetic body *V*. For this method to be exact, the discretisation cellsize must be chosen so as to divide the separate bodies $V_i$, as well as the empty space between them, into an integer number of cells in each dimension. For most cases of practical interest, this approach is not only restrictive in terms of the geometries that can be reasonably simulated, but also inefficient, since the resulting cellsize is typically much smaller than that required to accurately simulate each magnetic body separately. To give examples we distinguish two cases: i) magnetic multilayers with thickness values large compared to the separation between, and ii) ultrathin magnetic multilayers with relatively large separation between the layers. Case i) includes synthetic anti-ferromagnetic structures [14-17], whilst case ii) occurs most notably in ultrathin magnetic multilayered stacks used to study skyrmions [5-8,34]. With this method, the local and short-range effective field contributions, e.g. due to exchange interaction and magnetocrystalline anisotropy, are computed separately in each computational mesh, whilst the long-range demagnetising field is computed on the union of these computational meshes. We term this method supermesh convolution.

With the multilayered convolution approach, we can write the convolution sum as:

$$\mathbf{H}(\mathbf{r}'_{kl}) = -\sum_{\substack{i=1,\dots,n \\ \mathbf{r}_{ij}\in V_i}} \mathbf{N}(\mathbf{r}'_{kl} - \mathbf{r}_{ij}, \mathbf{h}_k, \mathbf{h}_i)\mathbf{M}(\mathbf{r}_{ij}), \quad k = 1,\dots,n; \quad \mathbf{r}'_{kl} \in V_k \tag{3}$$

In the demagnetising tensor of Equation **(**3**)** we explicitly specify the cellsize, **h**, of the two computational meshes the tensor relates. Since in Equation **(**3**)** we have *n* terms of the form given in Equation **(**1**)**, we can again apply the convolution theorem. This time for each output mesh (**H**) we have *n* input meshes (**M**), together with *n* kernels. Thus to calculate the outputs in all *n* meshes we require a total of $n^2$ sets of kernel multiplications and $n(n\text{-}1)$ summations in the transform space. This is illustrated in Figure 1. Since the set of *n* input magnetisation distributions is re-used for each of the *n* output field distributions, we only require *n* applications of the DFT, and similarly the final outputs can be obtained using only *n* applications of the inverse DFT. This approach is much more efficient than directly summing the inter-layer demagnetising field contributions, as this would require $n^2$ inverse DFTs.

In 3D mode (all DFTs are three-dimensional) we require all input/output spaces to have the same dimensions and same discretisation cellsize. For 2D mode (all DFTs are two-dimensional in the xy plane), there is no restriction on the thickness of each layer, thus we only require the x and y components of discretisation cellsizes to be the same across the *n* input/output spaces. In this latter case it is easy to modify the formulas given in Newell et al. [33] to calculate the tensor components for two cells with unequal dimensions – see Appendix A. If the *n* input spaces have unequal dimensions we can simply use zero padding to extend them to the largest dimension in the set. A more difficult case arises if the cellsize values differ between the input spaces. In this case we can use an interpolation method to first transfer the input values to scratch spaces with a common discretisation cellsize across the *n* scratch spaces. Similar remarks apply for the output fields, and this is also illustrated in Figure 1.

**Figure 1** – Multilayered convolution algorithm for *n* computational meshes. The magnetisation input of each mesh is transformed separately using a FFT algorithm, either directly (dotted line), or by first transferring to a scratch space with a common discretisation cellsize, using a weighted average smoother (solid lines). In the transform space the inputs are multiplied with pre-computed kernels for a total of $n^2$ sets of point-by-point multiplications. Finally the output demagnetising fields are obtained using an inverse FFT algorithm, which are set directly in the output meshes (dotted line), or transferred using a weighted average smoother if the discretisation cellsizes differ (solid lines).

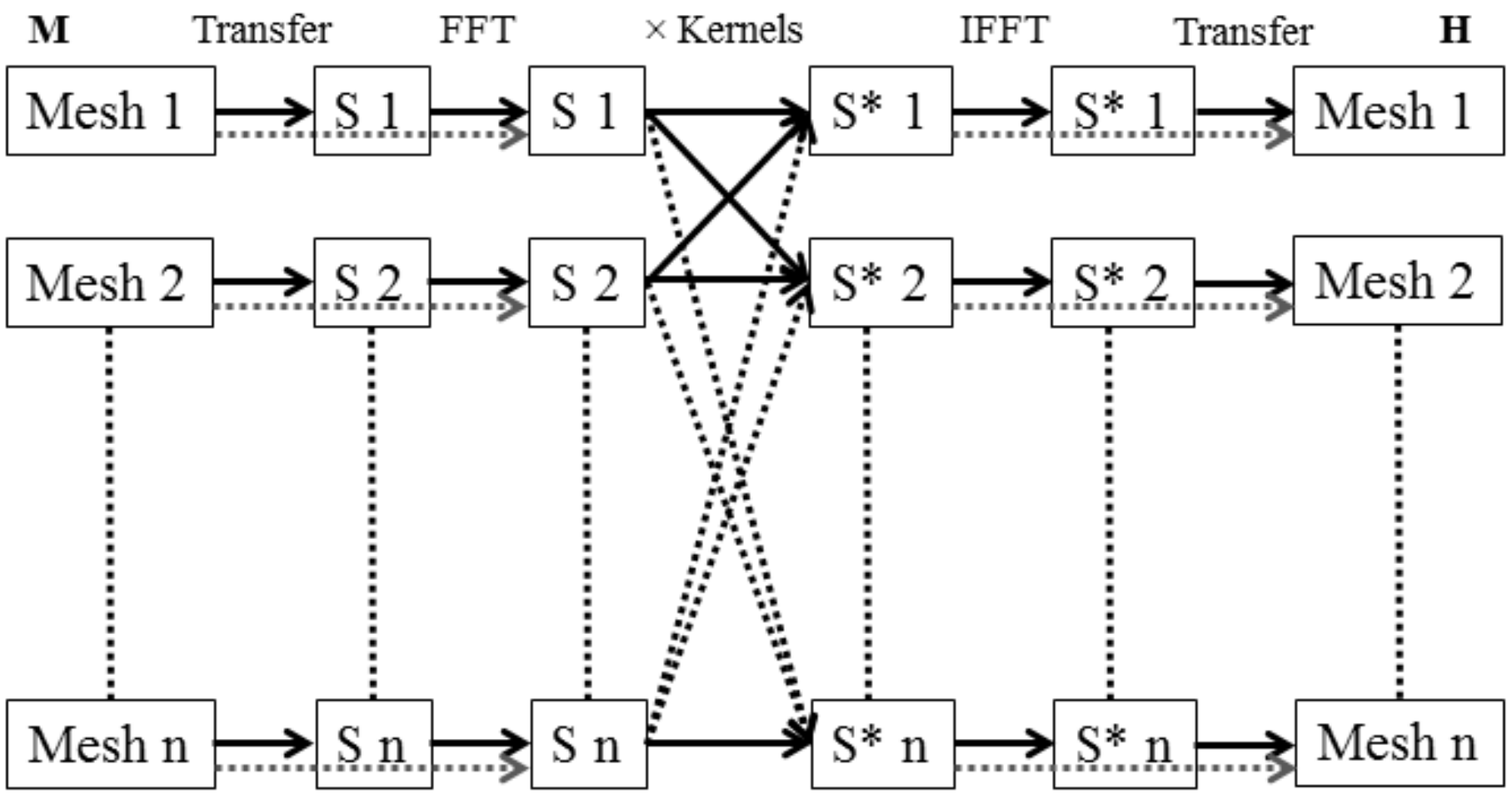

Before discussing the mesh transfer method, we note that in many micromagnetics problems involving multilayers, the simulated magnetic materials used are either the same, or with a similar exchange length, which means the required computational cellsize can be set the same without sacrificing computational efficiency. There is the further restriction on the cellsize due to the requirement of integer number of cells in each dimension. For 2D mode, as mentioned above, there is no restriction on the thickness of each layer as there is just one computational cellsize along the z direction for all computational meshes involved. In 3D mode we may also wish to simulate layers with different thickness values. In this case we can obtain the z component of the common discretisation cellsize by dividing the largest mesh z dimension by the largest number of computational cells along the z direction, from the set of *n* computational meshes. The input magnetisation distributions are then transferred using interpolation to the scratch spaces with common discretisation, using zero padding where needed. There is no restriction imposed on the cellsize by either the spacing or relative positioning between the layers, thus multilayered convolution allows for simulations with arbitrary spacing between the layers, which may be inaccessible to supermesh convolution. For example consider the Pt(5 nm)\FM\Au(*d*)\FM\Pt(5 nm) structure from Ref. [8], where FM is Ni(4 Å)\Co(7 Å)\Ni(4 Å), and *d* is a variable Au spacer thickness. To simulate such a structure using an exact discretisation, a 1 Å cellsize in the z direction is required which renders it impractical. Instead an effective medium approximation may be introduced by considering the FM layer as a whole, in which case a cellsize of 1.5nm can be used – this still requires discretising the Au layer which can be very inefficient for large thicknesses, and also restricts the values of *d* to multiples of 1.5nm. With the multilayered convolution method, this structure with the individual Ni and Co layers can be simulated exactly as is very efficiently (six 2D layers of the required thickness can be set), and moreover the Au layer thickness can be set to any value without impacting on the computational performance. Similar considerations apply to the multilayered structures used in Refs. [5-7], as well as the multilayered tracks used in Refs. [14-17]. Thus in many cases the individual layers may be simulated using 2D transforms, which further results in significant speedup compared to supermesh convolution, the latter requiring a large 3D convolution. The lower DFT dimensions also result in increased numerical accuracy [35]. The need for $n^2$ sets of kernel multiplications may seem excessive, but each set of point-by-point multiplications is much smaller compared to the case of supermesh convolution, which, when taking into account the significantly reduced DFT sizes, allows for a large number of layers to be handled while still providing significant speedup factors.

The mesh transfer procedure uses a weighted average smoother with second order accuracy in space [36], described as follows. Consider a discrete distribution of magnetisation values $\mathbf{M}$ at points $V = \{\mathbf{r}_i;\, i \in P\}$. Let $\mathbf{h}$ be the cellsize of the input mesh, with the set of cells $\{c_i;\, i \in P\}$ centered around the points $\mathbf{r}_i$. To obtain the magnetisation value at a point $\mathbf{r}'$ in a cell $c$ with dimensions $\mathbf{h}'$ we introduce the definitions $d_i = |\mathbf{r}' - \mathbf{r}_i|$, $d_V = |\mathbf{h}' + \mathbf{h}|/2$, and $\tilde{d}_i = d_V - d_i$. The weighted average is given as:

$$\mathbf{M}(\mathbf{r}') = \sum_{i \in P} w_i \mathbf{M}(\mathbf{r}_i), \quad \textbf{(4)}$$

where

$$
\begin{aligned}
& w_i = \frac{\tilde{d}_i \delta_i}{\tilde{d}_T} \\
& \delta_i = \begin{cases} 1, & c_i \cap c \neq \varnothing \\ 0, & otherwise \end{cases} \\
& \tilde{d}_T = \sum_{i \in P} \tilde{d}_i \delta_i
\end{aligned}
\quad \textbf{(5)}
$$

The weights can be pre-computed at the initialisation stage, thus speeding up the algorithm at run-time.

**Table 1** – Convolution kernel properties for the general 2D and 3D cases, as well as special cases, where $N_{dd}$ and $K_{dd}$ refer to the diagonal components (dd = xx, yy, or zz). 2D-Self and 3D-Self refer to calculation of self-demagnetising fields. 2D-zShift and 3D-zShift refer to cases where the shift between two computational meshes is along the z axis only. In general the storage space required has $(N_x/2+1)\times N_y \times N_z$ points. For *reduced* storage space (cases indicated in the table) we only need $(N_x/2+1)\times(N_y/2+1)\times(N_z/2+1)$ points. For 3D modes we require the cellsizes to match ($\mathbf{h}_i=\mathbf{h}_j$) for the computational meshes the kernel relates, whilst for 2D modes we only require the x and y components of the cellsizes to match ($h_{(x,y)i}=h_{(x,y)j}$). The symmetry properties of tensor components along the x, y, and z axes are indicated, as well as the resulting kernel types after DFT – real, imaginary, or complex.

| **Tensor** | **x** | **y** | **z** | **Kernel (DFT)** | **2D-Self** *reduced* | **3D-Self** *reduced* | **2D-zShift** *reduced* $h_{(x,y)i}=h_{(x,y)j}$ | **3D-zShift** *reduced* $\mathbf{h}_i=\mathbf{h}_j$ | **2D-Full** *full* $h_{(x,y)i}=h_{(x,y)j}$ | **3D-Full** *full* $\mathbf{h}_i=\mathbf{h}_j$ |
|---|---|---|---|---|---|---|---|---|---|---|
| $N_{dd}$ | even | even | even | $K_{dd}$ | real | real | real | complex | complex | complex |
| $N_{xy}$ | odd | odd | even | $K_{xy}$ | real | real | real | complex | complex | complex |
| $N_{xz}$ | odd | even | odd | $K_{xz}$ | 0 | real | imaginary | complex | complex | complex |
| $N_{yz}$ | even | odd | odd | $K_{yz}$ | 0 | real | imaginary | complex | complex | complex |

Finally we consider the properties of kernels used for transform space multiplications, which are obtained from the demagnetising tensors using the DFT. In general the kernels are complex-valued and use a storage space with $(N_x/2+1)\times N_y \times N_z$ points, where $N_x$, $N_y$, $N_z$ are the DFT sizes in the x, y, and z dimensions respectively. The first dimension is always reduced since the input tensor elements are purely real. The demagnetising tensor elements also have symmetry properties, which in some important special cases allow the kernels to be purely real or purely imaginary (thus resulting in multiplication by a scalar only), as well as use a reduced storage space of $(N_x/2+1)\times(N_y/2+1)\times(N_z/2+1)$ points [20]. The symmetry properties of the demagnetising tensor components, as well as the resulting kernel properties in the cases of interest are summarised in Table 1.

At one extreme we have the self-demagnetising kernels for 2D and 3D cases (i.e. zero shift between the input and output spaces), which, due to symmetry properties of the tensor components, are purely real and can also be stored using reduced storage space – the remaining elements may be recovered from symmetry properties of the kernels [20]. At the other extreme we obtain the stray field from one magnetic body at another, with an arbitrary shift between the two spaces. In this case the kernels are both complex-valued and require the full storage space. Whilst the input tensor elements have symmetries about the x = 0, y = 0, and z = 0 points in each dimension, due to the shift introduced the input tensor symmetries do not carry through to the transform space. The notable exception is that of a shift along the z axis only (cases denoted as 2D-zShift or 3D-zShift in Table 1). In this case the symmetries in the x and y dimensions are still applicable and the resulting kernel properties are summarised in Table 1. Note for the 3D-zShift case, whilst the kernels are complex they can still be stored using reduced storage space since the input to the z dimension DFT is either purely real or purely imaginary.

Whilst the multilayered convolution algorithm requires $n^2$ sets of kernel multiplications, typically we do not require storage of $n^2$ kernels due to redundant information between them. For example, for each kernel that relates an input and output space with a given shift between them, we also need a kernel for the opposite direction shift. For the 2D-zShift case this may simply be obtained from the first by adjusting signs in the kernel multiplication stage, as resulting from the tensor properties in Table 1. Also, since it is only the relative shift between two spaces that is important, not their absolute positions, we can further reduce the required kernel storage in many typical use cases. For example the most efficient use case is that of regularly spaced multilayers, for which we only need $n$ kernels. Finally, a note on implementation, the FFTs in Boris are computed using FFTW3 [37] on the CPU, and the CUDA 9.2 FFT library [38] on the GPU. Boris is coded in C++14 and is open source [32]. A pseudo-code for the multilayered convolution algorithm is shown in Appendix B.

## 3. Validation

To verify the multilayered convolution algorithm, micromagnetics problems have been solved using both the supermesh convolution and multilayered convolution algorithms for all the cases shown in Table 1. The most stringent test involves reproducing the exact magnetisation dynamics, similar to the approach taken in µMag standard problem 4 [39]. For these the Landau-Lifshitz-Gilbert (LLG) equation was solved with effective field contributions of applied field, exchange interaction field, and demagnetising field. An example of this is shown in Figure 2, where the magnetisation switching in a 640 nm × 320 nm trilayer $Ni_{80}Fe_{20}$ structure was simulated under a 20 kA/m in-plane magnetic field oriented 5° to the x axis. Typical material parameters for $Ni_{80}Fe_{20}$ are used as given in Ref. [40]. The starting magnetisation state is shown in the inset of Figure 2(b). The switching field was applied to the top and bottom layers only, thus the middle layer switches purely due to the dipole field from the outer layers. Here the outer layers have a thickness of 20 nm, whilst the middle layer has thickness of 10 nm, with separation between layers of 1 nm. The supermesh convolution method uses a (5 nm, 5 nm, 1 nm) cellsize in order to accommodate the 1 nm gap between the layers and was computed using mumax3. The Runge-Kutta 4$^{th}$ (RK4) order evaluation method was used with a 100 fs time step due to the stiffness of the LLG equation. For the multilayered convolution we simply use a 5 nm cubic cellsize in each of the three layers. This was computed using Boris with the RK4 evaluation method using a 500 fs time step. The magnetisation switching is plotted for the bottom and middle layers in Figure 2(b),(c) – the top layer average magnetisation dynamics is the same as for the bottom layer due to mirror symmetries. Due to the larger magnetic moments of the top and bottom layers, these are switched towards the applied magnetic field direction. The middle layer, with a smaller magnetic moment, switches due to the large stray fields from the top and bottom layers. As can be seen in Figure 2, the two convolution methods result in excellent agreement, despite the different cellsize values used to compute the demagnetising and exchange fields. This problem was computed using the GTX 1050 Ti GPU on Windows 7 x64. In terms of computational performance, the multilayered convolution on Boris is around 18 times faster compared to the supermesh convolution on mumax3, partly due to the much smaller time step required when using the smaller cellsize. In terms of absolute performance per RK4 iteration, the supermesh convolution on Boris is around 1.5 times faster compared to mumax3 on this platform.

**Figure 2** – Magnetisation switching in a trilayer $Ni_{80}Fe_{20}$ structure using a 20 kA/m in-plane magnetic field oriented 5° to the x axis, computed using multilayered convolution (Boris) as well as supermesh convolution (mumax$^3$). The top and bottom layers have thickness 20 nm, the middle layer has thickness 10 nm, with separation between layers of 1 nm, length of 640 nm, and width of 320 nm. (a) Magnetisation configuration during the switching event, showing the three separate layers, with magnetisation direction arrows color coded using the inset color wheel. (b), (c) Components of average magnetisation as a function of time, plotted for the bottom and middle layers respectively, showing the starting magnetisation configuration in the inset. For supermesh convolution a (5 nm, 5 nm, 1 nm) cellsize was used – open symbols – whilst for multilayered convolution a 5 nm cubic cellsize was used in each layer – dashed lines.

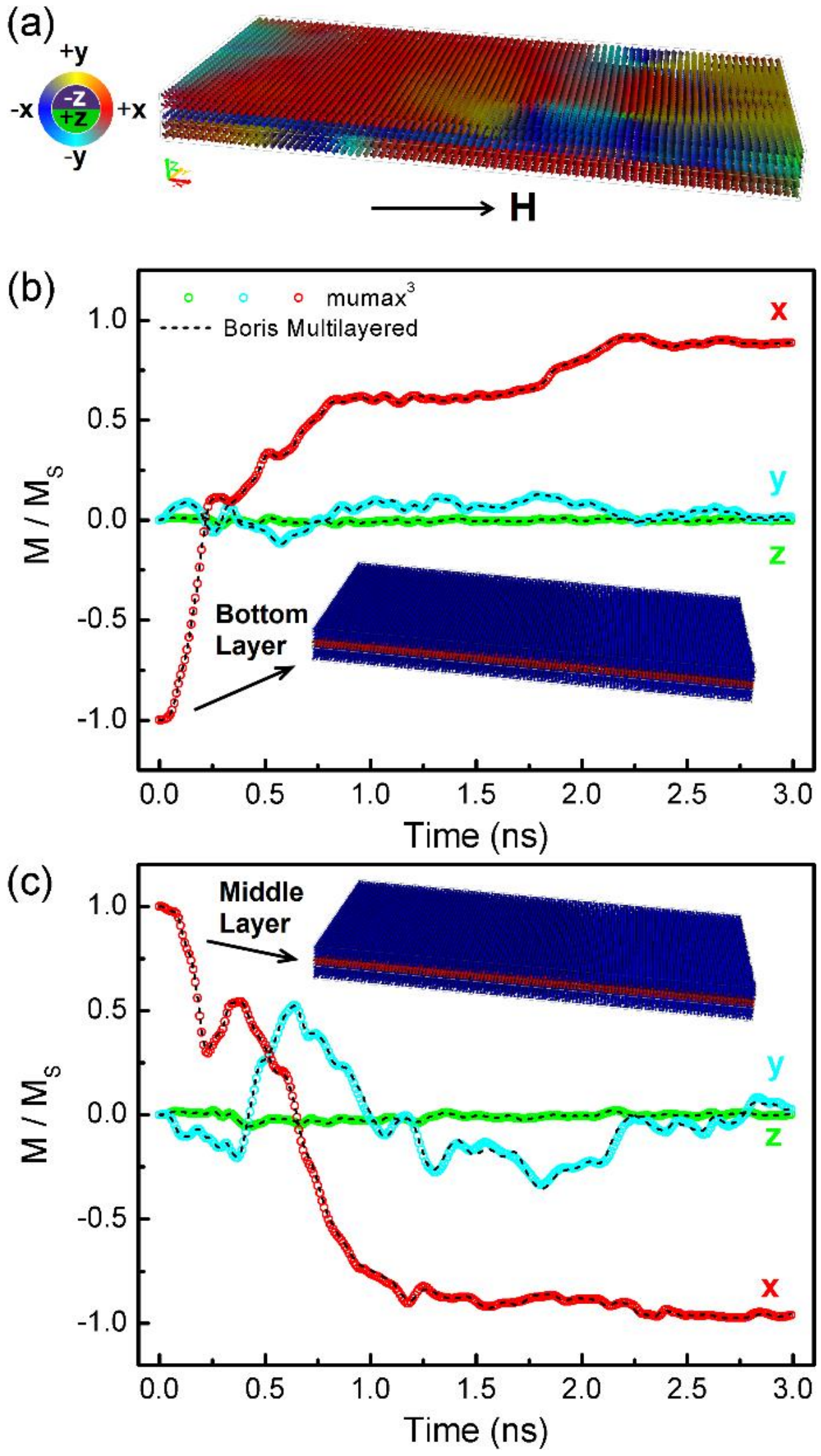

Another problem in shown in Figure 3, where the Néel skyrmion diameter in ultrathin Co layer stacks is computed using both the supermesh and multilayered convolution algorithms. The Co layers are 1 nm thick, of circular shape with 512 nm diameter, and with a 3 nm non-magnetic spacer between the layers. The Co layers have a strong perpendicular magnetocrystalline anisotropy, in practice arising due to interfacial spin-orbit coupling with a heavy metal layer [41], e.g. Pt, which forms part of the non-magnetic spacer. Material parameters used are the same as given in Ref. [34]. The effective field contributions include the applied field, exchange interaction field, interfacial DMI field with DMI exchange constant D = -1.5 mJ/m$^2$, uniaxial magnetocrystalline anisotropy field, and demagnetising field. The skyrmion diameter was obtained by fitting the z skyrmion profile with the function $m_z(r) = \cos(2\arctan(\sinh(R/w)/\sinh(r/w)))$ [42], where R is the skyrmion radius, and $w = \pi D/4K$ with $K = K_u - \mu_0 M_S^2/2$. Here $K_u$ is the uniaxial magnetocrystalline anisotropy and $M_S$ is the saturation magnetisation [34].

**Figure 3** – Calculation of average skyrmion diameter in multilayered 512 nm diameter disks as a function of out-of-plane magnetic field and number of Co layers. The Co layers are 1 nm thick with a separation of 3 nm. (a) Skyrmion in a 6-layer stack, with magnetisation direction arrows color coded using the inset color wheel. (b) Skyrmion diameter as a function of magnetic field and number of Co layers, computed with Boris using supermesh convolution (dashed lines), multilayered convolution (disks), as well as supermesh convolution with mumax3 (open squares).

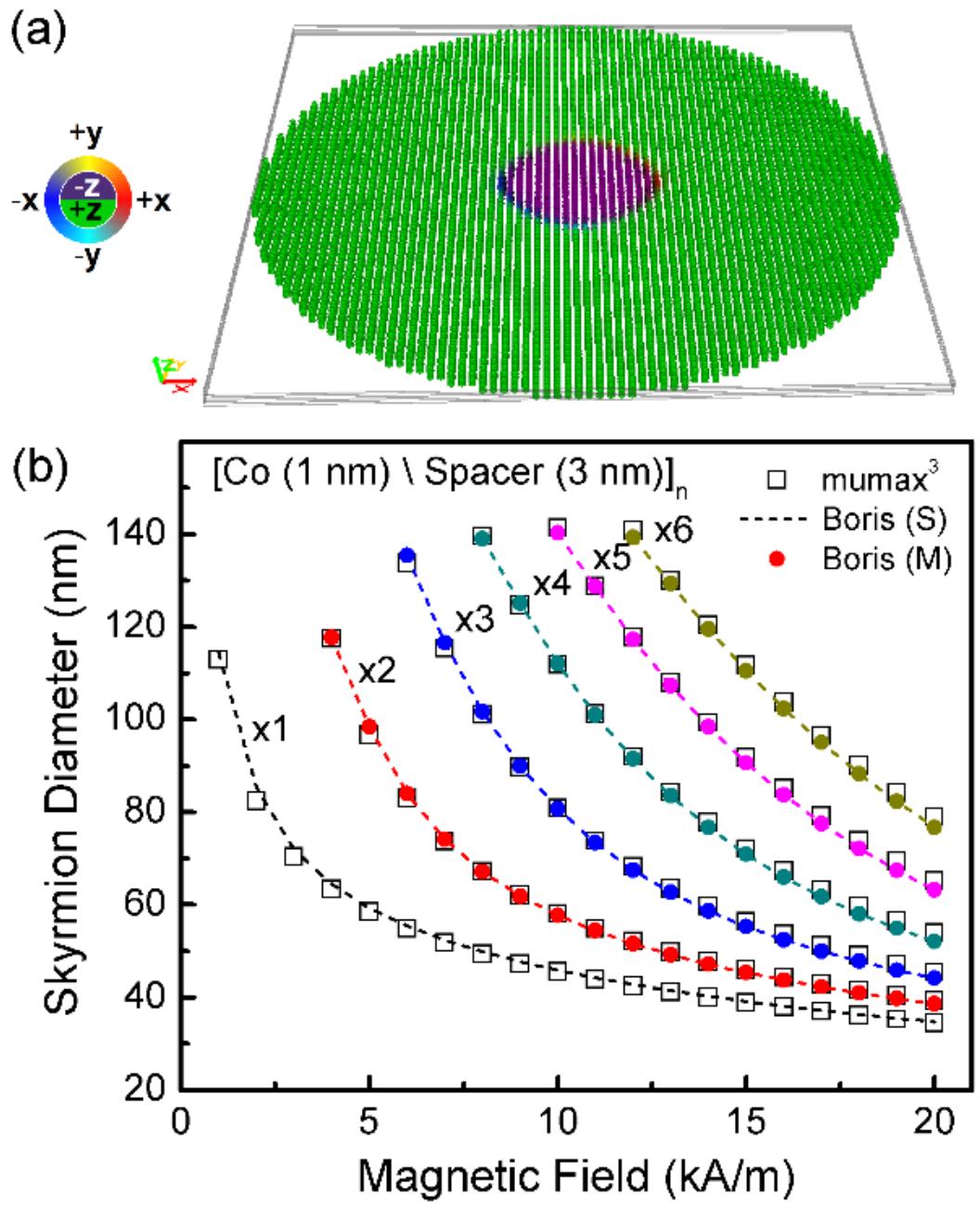

The calculated diameter as a function of out-of-plane magnetic field strength and number of Co layers is shown in Figure 3(b). Both the supermesh and multilayered convolution algorithms use a cellsize of (4 nm, 4 nm, 1 nm), however with multilayered convolution the stray field is only computed in the Co layers alone. As can be seen in Figure 3b, the computed diameters are virtually identical for the two methods on Boris, showing the expected inverse dependence on magnetic field strength [43]. We have also computed the skyrmion diameters using supermesh convolution with mumax3, shown as open squares in Figure 3b. Again there is an excellent agreement between the two methods, with differences in diameter up to 2 nm, thus half the in-plane discretisation cellsize.

## 4. Algorithm Performance

The performance comparison between the two algorithms clearly depends on the relative spacing between the layers. At one extreme we can have a set of magnetic layers with relatively little empty space between them, and which can also be exactly discretised without reducing the cellsize dimensions just to accommodate the layer spacing. In this case the supermesh convolution algorithm is faster. At the other extreme we have magnetic multilayers with either very small spacing between them relative to the layer thickness values, or which otherwise need a very small magnetic cellsize to exactly and uniformly discretise for supermesh convolution. Here we give a performance comparison for a typical use case, e.g. as arising in multilayered [Pt (3 nm)\Co (1 nm)\Ta (4 nm)]$_n$ stacks used in previous works [5,34]. The same material parameters and effective fields are used as for the results in Figure 3. Two computational platforms were used, GTX 980 Ti GPU with the i7 4790K CPU on Windows 7 x64, as well as the GTX 1050 Ti GPU with the i7 x980 CPU on Windows 10 x64. The benchmarking results are the average of the results obtained on these 2 computational platforms. The benchmarking results are shown in Figure 4, plotting the computation time per iteration as a function of number of stack repetitions, namely $n$ = 1 up to $n$ = 17, for both the supermesh and multilayered convolution algorithms. Both the CPU and GPU implementations of the algorithms are considered. In all cases the multilayered convolution algorithm results in much faster performance, with speedup factors between 2.5 and 8. The multilayered convolution simulation time increases smoothly with number of layers, following a parabolic dependence partly due to the required $n^2$ sets of kernel multiplications indicated in Figure 1. On the other hand the supermesh convolution algorithm shows abrupt jumps in simulation time – this is due to the power-of-2 dimensions required by the FFT algorithm. The same benchmarking test was run for the case of 1 nm separation between the layers, which should favour supermesh convolution due to the reduced empty space between the layers. Even in this case the multilayered convolution algorithm is faster, with an average speedup factor of 1.5, and a maximum speedup factor of 2. The only case where multilayered convolution was slower was for *n* = 16, with a speedup factor of 0.9, however this jumps to 1.2 for *n* = 17 as the z FFT dimension is doubled for supermesh convolution.

**Figure 4** – Performance comparison of supermesh and multilayered convolution algorithms in a Co stack as a function of number of Co repetitions, for both (a) CPU implementation for 512 nm diameter disks, and (b) GPU implementation for 1024 nm diameter disks. Solid disks show the simulation time per iteration, and empty triangles show the speedup factor of multilayered versus supermesh convolution (simulation time ratio of supermesh to multilayered convolution).

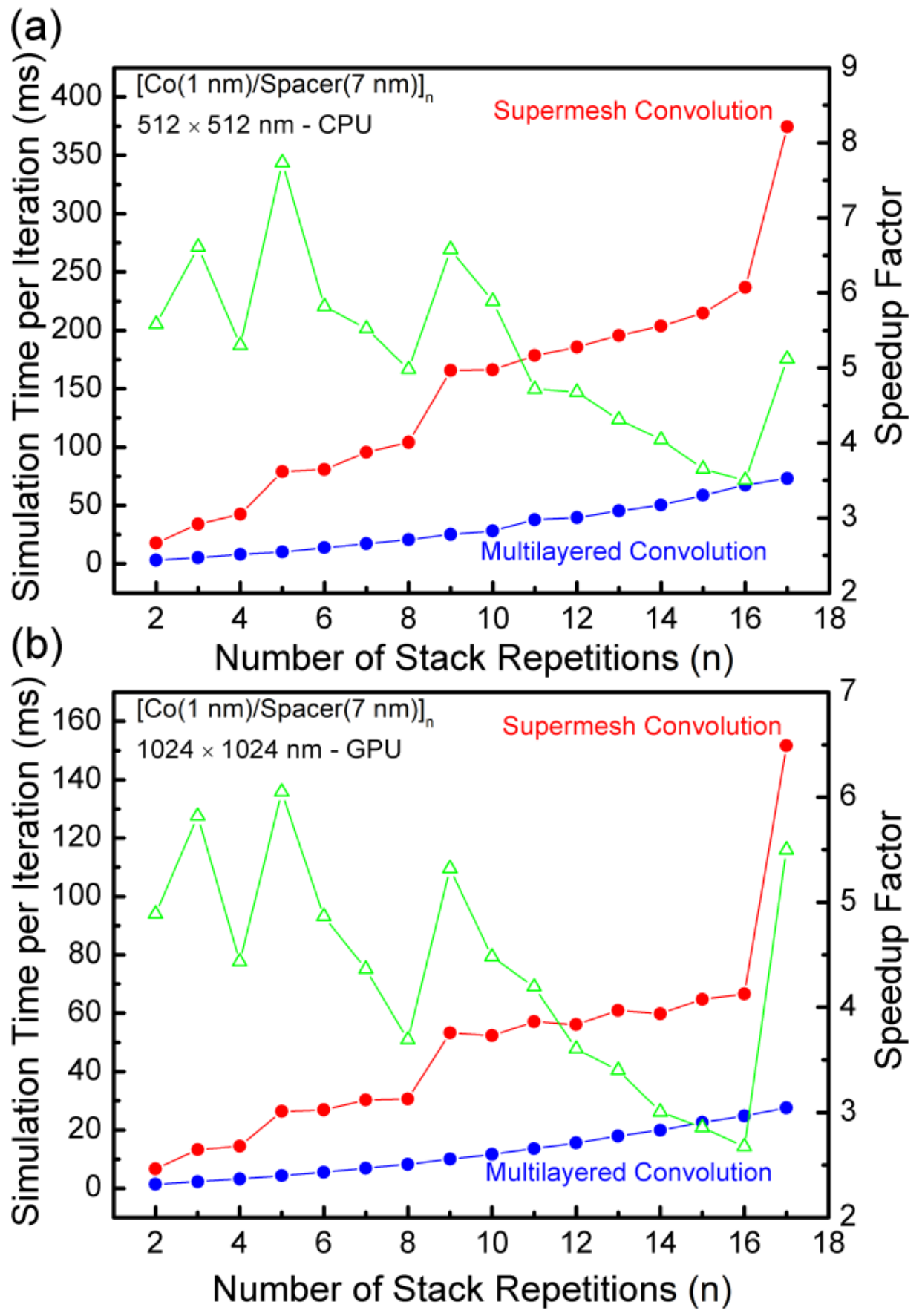

## 5. Conclusions

Here we have demonstrated a new method of computing demagnetising fields in magnetic multilayers, which was shown to be a generalisation of the FFT-based convolution method used for single magnetic bodies. The multilayered convolution method is able to handle arbitrary spacing and arbitrary relative positioning between the magnetic layers with no impact on the computational performance. Moreover, for thin magnetic layers, which may be simulated using 2D convolution, the multilayered convolution method also allows arbitrary thickness values for the layers in the stack. For 3D convolution the algorithm is also able to handle layers with different thickness values, as well as different xy plane dimensions between the different layers. The algorithm was implemented both for the CPU and GPU. Multilayered convolution is most efficient when the individual layers are thin and are stacked along the z direction. This case occurs very often in practice, and in particular for a typical multilayered stack used to study skyrmions it was shown to be up to 8 times faster compared to the simple convolution method which treats the entire multilayered stack as a single magnetic body.

## Appendix A

Let $\mathbf{s} = (x, y, z)$ be the shift between two rectangular prisms with dimensions (cellsizes) $\mathbf{h_s} = (h_x, h_y, h_{sz})$, and $\mathbf{h_d} = (h_x, h_y, h_{dz})$ respectively; thus the cellsizes are allowed to differ at most in their z dimension. The shift is oriented from the origin corner of the source cellsize with dimensions $\mathbf{h_s}$ to the origin corner of the destination cellsize with dimensions $\mathbf{h_d}$. In this case the demagnetising tensors for the *xx* and *xy* elements are computed using:

$$\begin{aligned} N_{xx}(\mathbf{s}) &= L[f;\mathbf{h_s},\mathbf{h_d}](\mathbf{s}) \\ N_{xy}(\mathbf{s}) &= L[g;\mathbf{h_s},\mathbf{h_d}](\mathbf{s}) \end{aligned} \qquad \mathbf{(6)}$$

The function *L* is given as:

$$L[\phi;\mathbf{h_s},\mathbf{h_d}](\mathbf{s}) = \frac{1}{\tau}\sum_{\varepsilon_1,\varepsilon_2=-1}^{1}\frac{1}{-2^{|\varepsilon_1|+|\varepsilon_2|}}\begin{Bmatrix} -\phi(x+\varepsilon_1 h_x, y+\varepsilon_2 h_y, z-h_{sz}) \\ -\phi(x+\varepsilon_1 h_x, y+\varepsilon_2 h_y, z+h_{dz}) \\ +\phi(x+\varepsilon_1 h_x, y+\varepsilon_2 h_y, z) \\ +\phi(x+\varepsilon_1 h_x, y+\varepsilon_2 h_y, z-\Delta) \end{Bmatrix}, \qquad \mathbf{(7)}$$

where $\tau = \pi h_x \times h_y \times h_{dz}$, and $\Delta = h_{sz} - h_{dz}$.

The functions *f* and *g* are given below [33], where $R^2 = x^2 + y^2 + z^2$.

$$\begin{aligned} f(x,y,z) = {} & \frac{(2x^2-y^2-z^2)R}{6} - xyz\arctan\left(\frac{yz}{xR}\right) \\ & + \frac{y(z^2-x^2)}{4}\ln\left(1+\frac{2y(y+R)}{x^2+z^2}\right) + \frac{z(y^2-x^2)}{4}\ln\left(1+\frac{2z(z+R)}{x^2+y^2}\right) \end{aligned} \qquad \mathbf{(8)}$$

$$\begin{aligned} g(x,y,z) = {} & -\frac{xyR}{3} - \frac{z^3}{6}\arctan\left(\frac{xy}{zR}\right) - \frac{zy^2}{2}\arctan\left(\frac{xz}{yR}\right) - \frac{zx^2}{2}\arctan\left(\frac{yz}{xR}\right) \\ & + \frac{y(3z^2-y^2)}{12}\ln\left(1+\frac{2x(x+R)}{y^2+z^2}\right) + \frac{x(3z^2-x^2)}{12}\ln\left(1+\frac{2y(y+R)}{x^2+z^2}\right) \\ & + \frac{xyz}{2}\ln\left(1+\frac{2z(z+R)}{x^2+y^2}\right) \end{aligned} \qquad (9)$$

The remaining tensor elements may be obtained from $N_{xx}$ and $N_{xy}$ by permuting the dimensions for the **s**, $\mathbf{h_s}$, and $\mathbf{h_d}$ vectors as explained in Ref. [33].

## Appendix B

The multilayered convolution algorithm is presented in pseudo-code below. The implementation using C++, both for the CPU and GPU using CUDA, is available as open source in Ref. [32].

**DATA**:
Mesh $M_1, \ldots, M_n$.
Scratch $S_1, \ldots, S_n, S_1^*, \ldots, S_n^*$
Each Mesh and Scratch space has a *rectangle* and *cellsize*
Each Mesh has *magnetisation* and *field* data
Kernel $K_{ij}$ for $i,j = 1, \ldots, n$

**INITIALIZATION**:
**for** $i$ in range 1, …, n **do**
- Set $S_i$ and $S_i^*$ rectangle origin the same as that of $M_i$
- Set $S_i$ and $S_i^*$ rectangle dimensions as the largest dimensions in the set of mesh rectangles $\{M_1, \ldots, M_n\}$
- Set $S_i$ and $S_i^*$ cellsizes as the ratio of their rectangles to the maximum number of computational cells from $\{M_1, \ldots, M_n\}$ in each dimension respectively

**end for**
**for** $i$ in range 1, …, n **do**
    **for** $j$ in range 1, …, n **do**
        $T_{ij}$ = Compute_Newell_Tensor_Elements($S_i$, $S_j$)
        $K_{ij}$ = FFT($T_{ij}$)
    **end for**
**end for**

**procedure** Multiconvolution($M_i, \ldots, M_n, S_1, \ldots, S_n, S_1^*, \ldots, S_n^*$)

    **for** $i$ in range 1, …, n **do**
        **if** $M_i$.*rectangle* = $S_i$.*rectangle* **and** $M_i$.*cellsize* = $S_i$.*cellsize* **then**
            $S_i$ = FFT($M_i$.*magnetisation*)
        **else**
            $S_i$ = Transfer($M_i$.*magnetisation*)
            $S_i$ = FFT($S_i$)
        **end if**
    **end for**

    **for** $i$ in range 1, …, n **do**
        **for** $j$ in range 1, …, n **do**
            $S_i^* \mathrel{+}= S_j \times K_j$, using point-by-point multiplication
        **end for**
    **end for**

    **for** $i$ in range 1, …, n **do**
        **if** $M_i$.*rectangle* = $S_i^*$.*rectangle* **and** $M_i$.*cellsize* = $S_i^*$.*cellsize* **then**
            $M_i$.*field* = Inverse_FFT($S_i^*$)
        **else**
            $S_i^*$ = Inverse_FFT($S_i^*$)
            $M_i$.*field* = Transfer ($S_i^*$)
        **end if**
    **end for**
**end procedure**